%% file: KDD_main.tex
\pdfoutput=1
\documentclass[sigconf]{acmart}

\usepackage{booktabs} 
\usepackage{amsmath}
\usepackage{balance}       
\usepackage{graphics}      
\usepackage[T1]{fontenc}   
\usepackage{color}
\usepackage{booktabs}
\usepackage{textcomp}
\usepackage{tikz}
\usetikzlibrary{bayesnet}
\usepackage{mathtools}

\setcopyright{acmcopyright}

\acmDOI{10.475/123_4}

\acmISBN{123-4567-24-567/08/06}

\acmConference[KDD'17]{Machine Learning for Education (ml4ed) Workshop at ACM Knowledge Discovery and Data Mining Conference}{August 2017}{Halifax, Nova Scotia, Canada} 
\acmYear{2017}
\copyrightyear{2017}

\acmPrice{15.00}

\begin{document}
\title{STEM-ming the Tide: Predicting STEM attrition using student transcript data}

\author{Lovenoor Aulck, Rohan Aras, Lysia Li, Coulter L'Heureux, Peter Lu, Jevin D. West}
\authornote{Aras, Li, L'Heureux, and Lu contributed equally to this work.}
\affiliation{%
  \institution{University of Washington Information School}
  \city{Seattle} 
  \state{Washington} 
}
\email{laulck, rohana, ctl14, lysialee, lupeter5, jevinw@uw.edu}

\renewcommand{\shortauthors}{L. Aulck et al.}

\begin{abstract}
Science, technology, engineering, and math (STEM) fields play growing roles in national and international economies by driving innovation and generating high salary jobs. Yet, the US is lagging behind other highly industrialized nations in terms of STEM education and training. Furthermore, many economic forecasts predict a rising shortage of domestic STEM-trained professions in the US for years to come. One potential solution to this deficit is to decrease the rates at which students leave STEM-related fields in higher education, as currently over half of all students intending to graduate with a STEM degree eventually attrite. However, little quantitative research at scale has looked at causes of STEM attrition, let alone the use of machine learning to examine how well this phenomenon can be predicted. In this paper, we detail our efforts to model and predict dropout from STEM fields using one of the largest known datasets used for research on students at a traditional campus setting. Our results suggest that attrition from STEM fields can be accurately predicted with data that is routinely collected at universities using only information on students' first academic year. We also propose a method to model student STEM intentions for each academic term to better understand the timing of STEM attrition events. We believe these results show great promise in using machine learning to improve STEM retention in traditional and non-traditional campus settings.

\end{abstract}

%
%



\maketitle

\input{kdd_body}

\bibliographystyle{ACM-Reference-Format}
\bibliography{bestcapstone} 

\end{document}

%% file: kdd_body.tex
\section{Introduction}
\subsection{Background}
The United States' (US') standing in the twenty-first-century global economy will largely depend on its ability to foster a workforce well-equipped with skills in fields of science, technology, engineering, and math (STEM) \cite{national2016developing, augustine2010rising}. This is because the growing pervasiveness and ubiquity of technology in modern society has increased STEM fields' central roles in driving economic growth and producing jobs through innovation and development \cite{rothwell2013hidden}. That said, by many measures, the US is lagging behind other world leaders in training students in STEM fields \cite{savkar2010time, beach2013us, kelly2013performance} and must increase the number of students graduating with STEM degrees by about 33\% just to fill future needs in the domestic workforce \cite{pcast2012engage}. Considering the importance of STEM-driven industries to national and regional economies, these impending shortages of STEM-trained professionals have brought calls for systemic countermeasures in the US \cite{national2016developing, augustine2010rising, pcast2012engage, graham2013increasing, mervis2010better}. One potential solution to these shortages is decreasing the attrition/dropout of students from STEM fields. In fact, it is believed that 50-60\% of students entering US colleges intending to major in a STEM field ultimately either graduate with a non-STEM degree or do not graduate at all \cite{pcast2012engage, gates2012engage, chen2013stem, chen2015stem}. In many cases, these students leaving STEM fields are capable students who could have made valuable additions to the STEM workforce \cite{chen2015stem, bettinger2010or, lowell2009steady}. A 2012 report to the White House stated that decreasing the national yearly STEM attrition rate by just 10 percentage points from 2012 onwards will help reduce the impending shortage of STEM professionals in 2022 by 75\% \cite{pcast2012engage, gates2012engage}.

Thus far, little has been done to analyze and understand STEM attrition from a data-/machine learning-centric perspective. In this work, we model STEM attrition using data gathered from the registrar databases of a large, publicly-funded US university. The broader objective of this work is to better understand key determinants of student attrition from STEM fields while also developing methods to identify which students are at risk of transitioning away from STEM fields during the course of their undergraduate academics. Ultimately, we believe this work can influence policy decisions and provide recommendations for interventions to improve STEM-related student outcomes. In this paper, we present results from our attempts to use student-level transcript data to: 1) model and predict eventual STEM graduation using only students' first-year academic records and 2) identify and predict the time at which students transition away from STEM fields during their undergraduate academic careers. In this sense, we are not only interested in predicting \textit{whether} students will attrite from STEM tracks but also \textit{when} this occurs. We believe this to be the first look at STEM attrition at a large university from a machine learning (ML) perspective; furthermore, the size of the dataset used in this study (over 24,000 students) is believed to be among the largest used for an attrition study in a traditional university campus setting, STEM-specific or otherwise.

\subsection{Related Work}
This work primarily relates to two bodies of literature: studies on STEM attrition and studies on predictive modelling using student data. The former has mostly centered on surveys or regression analysis of longitudinal data. For example, Chen and Soldner used longitudinal data to examine student attrition from STEM fields between 2004-09, finding STEM retention to be correlated with a wide range of factors, including demographic and academic variables \cite{chen2013stem}. In later work, Chen examined the causes for STEM attrition among high-performing college students using the same data source, noting that the intensity of students' early STEM coursework has a large influence on their decision to leave STEM tracks of study \cite{chen2015stem}. Meanwhile, Rask used transcript records of 5,000 students at a small liberal arts college to find that grades in early science courses as well as pre-college preparedness (as measured by standardized test scores) greatly influence STEM retention \cite{rask2010attrition}. Bettinger looked at students who entered Ohio colleges in the late 1990s to find that earnings potential often drives high-performing students' decisions to move away from STEM fields \cite{bettinger2010or}. Beyond this, there is still a need for further exploration of the extent and causes of undergraduate STEM attrition, as evidenced by recent conferences with education leaders\footnote{\url{http://usnewsstemsolutions.com/}}.

The literature on predictive modelling using student data, meanwhile, is relatively recent and based on principles in educational data mining (EDM) - a field centering on gaining insights from what are typically large sets of educational data \cite{romero2010educational}. Much research on attrition and dropout using EDM has centered on massive online open courses (MOOCs) and other online environments \cite{lykourentzou2009dropout,yang2013turn,halawa2014dropout}. However, studying attrition in MOOCs and other online settings lends itself to much more expansive data collection opportunities and a more detailed monitoring of students, as emphasized by Niemi \cite{niemi2012using}, thereby limiting the extent to which this work can be generalized to more traditional campus settings. Meanwhile, EDM-centric work on attrition in more traditional campus settings has been rather scarce and usually limited to small, homogeneous subsets of students. As examples, Moseley predicted the attrition/graduation of 528 nursing students using rule induction methods without controlling for the amount of information available for each student (i.e. the number of terms/semesters examined) \cite{moseley2008predicting}. Dekker et al. looked at only the first semester grades of 648 students in the Electrical Engineering department at the Eindhoven University of Technology and predicted dropout with relatively strong ($>$75\%) accuracy \cite{dekker2009predicting}. Kova{\v{c}}i{\'c} used tree-based methods on a similarly-sized dataset of 453 students at the Open Polytechnic of New Zealand, finding ethnicity and students' course taking patterns to be highly useful in prediction \cite{kovavcic2010early}. Studies with similarly-sized cohorts also focused on informatics or engineering students at Masaryk University \cite{bayer2012predicting}, the University of Notre Dame \cite{aguiar2014engagement}, and an undisclosed public university in the US Midwest \cite{lin2009student}. These and similar studies, however, tend to focus on relatively small groups of students (typically $<$2,000) with similar academic interests. What's more, no previous study has focused specifically on predicting student attrition from across the spectrum of \textit{STEM} fields.

The work we present more closely relates to a subset of literature looking at student attrition in the context of the heterogeneity of students across an entire campus and not just a subset thereof. Delen used 8 years of institutional data on over 6,000 students (after accounting for imbalanced classes) at a large, public US university to predict whether the students would return for their second year \cite{delen2011predicting}. Ram et al. used data on about 6,500 freshmen at a large, public US university to predict whether students would drop out after their first semester, and for those that remained, whether they will drop out after their second semester \cite{ram2015using}. Our group has also shown some early success in predicting dropout in more heterogeneous data while using a much larger dataset ($>$32,000 students) \cite{Aulck2016}.

In contrast to the above, this work seeks to predict attrition specifically from STEM fields while using a dataset composed of an extremely heterogeneous population across all undergraduates at a large, public US university who intend on pursuing a STEM degree. We also propose a proof-of-concept approach to trace student trajectories into and away from STEM fields during their undergraduate academics, which we hope can help provide further clarity on the STEM attrition phenomena and inspire real-time intervention. In all, we believe this work holds great promise to be expanded for use in policy decisions and university student advising.

\section{Methods}

\subsection{Data}
\label{subsec:data}
weDe-identified, psuedonymized data were gathered from the University of Washington's (UW) registrar databases in early 2017. The process of obtaining approval for, gathering, and curating this data took over 2 years to complete. The data contain the demographic information (race, ethnicity, gender, birth date, and resident status), pre-college entry information (standardized test scores, high school grades, parents' educational attainment, and application zip code, where available), and complete transcript records (courses taken, when they were taken, grades received, and majors declared) for all students in the University of Washington (UW) system. We did not have access to financial aid information. For this work, we focused on matriculated undergraduate students at UW's main campus in Seattle, WA who first enrolled between 1998 and 2010 as freshmen entrants (i.e. those who did not transfer from another post-secondary institution, excluding running start and college in high school students). The year 2010 was used as a cutoff for the data to allow students first enrolling in 2010 to have 6 full years to graduate from the time of first enrollment. In all, this was 66,060 students. 

To focus on the issue of STEM attrition, we looked at students'  majors during their first calendar year on campus. Specifically, a majority of freshmen students enter the UW system without a major designation and are not required to formally declare a major until the end of their 2nd year. As such, most students enter UW with a ``pre-major'' designation of some sort. To predict STEM attrition, only students who had either declared a major in their first year that was classified as a STEM major by UW\footnote{UW's STEM designations are based on the federal and state classification of Classification of Instructional Programs (CIP) codes.} or who had a pre-major designation that corresponded to a STEM line of study (e.g. pre-engineering, pre-health sciences, pre-physical sciences, etc.) were included. Some UW majors have multiple degree tracks and for two of these majors (math and design), some tracks were considered STEM while some were not. The math degree had four tracks, one of which was a non-STEM teaching preparation track; similarly, the design degree also had four tracks, of which one (interaction design) was considered STEM while the others were not. The data in the registrar's student major and degree database, however, did not always distinguish which degree track within a major a student pursued. As such, any degree that had at least half of its tracks considered STEM was considered a STEM degree as a whole - in the above cases, math was considered a STEM degree and design was not. 

The dataset of students intending to major in STEM fields included 24,341 students and are henceforth referred to as ``STEM students'' in reference to their desire to obtain STEM degrees. The demographics of the student sample along with graduation rates in STEM (i.e. STEM persistence rates) are shown in Table \ref{tab:demo}.

\begin{table}
  \caption{Demographics of STEM students}
  \label{tab:demo}
  \begin{tabular}{lcc|c}
    \toprule
    & STEM Grads & STEM NCs & \shortstack{STEM\\Grad Rate}\\
    \midrule
    All & 12,121 & 12,220 & 49.80\%\\
    \midrule
    \hspace{0.5cm} \textsc{Gender}  & & & \\
    Female & 4,446 & 5,951 & 42.76\%\\
    Male & 7,659 & 6,257 & 55.02\%\\
    Other/Unknown & 16 & 12 & 57.14\%\\
    \midrule
    \hspace{0.5cm} \textsc{Race}  & & & \\
    African Amer. & 177 & 460 & 27.79\%\\
    Amer. Ind. & 106 & 236 & 30.99\%\\
    Asian & 4,059 & 3,857 & 51.28\%\\
    Caucasian & 6,196 & 6,149 & 50.19\%\\
    Hawaiian/Pac. Is. & 76 & 123 & 38.19\%\\
    Other/Unknown & 1,507 & 1,396 & 51.93\%\\
    \midrule
    \hspace{0.5cm} \textsc{Ethnicity}  & & & \\
    Hispanic & 419 & 813 & 34.01\%\\
    Not Hispanic & 11,702 & 11,407 & 50.64\%\\
    \midrule
    \hspace{0.5cm} \textsc{Residency}  & & & \\
    State Resident & 9,633 & 9,745 & 49.71\%\\
    Not Resident & 2,488 & 2,475 & 50.13\%\\
  \bottomrule
\end{tabular}
\end{table}

\subsection{Defining STEM graduates}
\label{subsec:grads}
``Graduates'' (grads) were defined as students who completed at least one undergraduate degree within 6 calendar years of first enrollment at UW; ``non-completions'' (NCs) were defined as all students who did not graduate. UW uses a quarter-based academic calendar and a 6-year time-to-completion was treated as 24 calendar quarters after the quarter of first enrollment. Enrollment was defined as when a student received at least one transcript grade (regardless of whether it is numeric or passing) as a matriculated student for a term. The overall graduation rate in the dataset was about 78.5\% based on these definitions of graduate and non-completion. 

STEM graduates were STEM students who met the above criteria for being a graduate and graduated with a degree in a major that was classified as STEM by UW.  Meanwhile, any student who was designated a STEM student as outlined in Section \ref{subsec:data} but did not ultimately graduate with a STEM degree was labelled as a STEM NC. It should be noted that STEM NCs include STEM students who graduated with non-STEM degrees but not with STEM degrees. In the dataset, 12,121 of 24,341 STEM students graduated with a STEM degree (49.80\%) while 50.20\% of students either did not graduate or graduated with a non-STEM degree. This indication of whether or not a STEM student was a STEM graduate was used as the binary outcome variable for predicting STEM attrition. The nearly even split of students into the two classes (STEM graduates and STEM NCs) obviated the need for any balancing of the classes.

\subsection{Predicting STEM attrition}
\label{subsec:predictions}
\subsubsection{Feature engineering}
\label{subsubsec:features}
In predicting students' graduation with a STEM degree, only student data with information prior to students' attending UW and transcript records up through one calendar year from their first enrollment (i.e. four consecutive academic quarters from their first enrollment) at UW was used for modelling. First-year transcript data for the STEM students consisted of 270,833 individual records of courses taken. The information on students prior to their entry into UW included demographic information and pre-entry information as described in Section \ref{subsec:data}. In addition, US census data was used to add additional information based on students' application ZIP code including: average income in ZIP, percentage of high school graduates in ZIP, and percentage of college graduates in ZIP. In addition, we also calculated the distance from the ZIP code to UW using Python's geopy package\footnote{\url{https://geopy.readthedocs.io/en/1.10.0/}}. 

Most demographic variables were categorical variables where each student only belonged to a single category with mutually exclusive inclusion in categories across variables. Each possible race, ethnicity, gender, and resident status were mapped across dummy variables. SAT and ACT scores, meanwhile, were available for 94\% and 27\% of the data, respectively. To impute missing SAT and ACT scores, we used a linear regression model with other demographic and pre-college entry data. In our previous work, we had tried using mean imputation for these missing values and did not see an appreciable difference in performance \cite{Aulck2016}. Students' pre-entry and demographic information accounted for 52 total features.

The transcript data consisted of individual entries for each course taken by each student, which included the students' grade, declared major at the time the course was taken, and when the course was taken. For each student grade, in addition to the institutionally-used GPA, a percentile score and z-score were also calculated. In the below descriptions of features, we use student ``performance'' to refer to five measures: counts of the number of courses taken, the number of credits earned, average GPA, average z-score, and average percentile grade. We calculated measures of student performance across the first year. In addition, we also calculated measures of students' performance in their respective first, second, third, and fourth quarters individually as well as student performance in the last academic quarter for which they were enrolled in their first year. Differences between students' performance in successive quarters were also calculated as were counts of the number of failed courses and withdrawals. In addition, measures of student course sizes and grade distributions in courses were also calculated. These first-year summary measures accounted for 116 features. Additionally, each possible major declared by students was given a separate feature (150 total).

In addition to summary data across the first year, we also calculated student performance in specific course groupings. This included performance in 100-, 200-, 300-, and 400-level courses as well as performance in remedial (sub-100-level) courses. In addition, student performance in STEM ``gatekeeper'' courses was also calculated, which included UW's intro-level, year-long calculus, biology, chemistry (inorganic and organic grouped together), and physics series. This accounted for 60 additional features. Finally, we also calculated student performance in courses for each departmental prefix for which at least 6 students in the dataset had taken a course. This accounted for another 1000 features across 200 course prefixes. In total, there were 1,378 features for each student.

\subsubsection{Machine learning experiments}
We report results from four ML models to predict the binary STEM dropout variable outlined in Section \ref{subsec:grads} using the features described in Section \ref{subsubsec:features}. The four models used were: regularized logistic regression, random forests, adaptive boosted logistic regression (ada-boosted log reg), and gradient boosted decision trees (gradient-boosted trees). In these supervised ML experiments, we use a 80\% random sample of STEM students (training set) to tune model hyperparameters (e.g. the regularization strength for logistic regression and the depth of the trees in random forests) using 10-fold cross-validation. Measures of performance are reported on the remaining 20\% of data (4,862 observations; test set), which is not used in cross-validation and hyperparameter tuning.

To better understand individual factors related to STEM graduation, we also run 1,378 individual logistic regressions of graduation using each feature in the dataset in isolation. These models are trained on the same training observations (i.e. students) and performance is reported on the same test observations as above.

\subsection{Analyzing STEM affinities}
\label{subsec:affinities}
The methods detailed in previous sections use supervised ML approaches to predict whether students shift from STEM-based educational pursuits. As stated earlier, while understanding factors related to whether students graduate with STEM degrees is of great interest, we are also interested in when students transition away from STEM fields. To better understand the timing of these transitions away from STEM, we present a proof-of-concept probabilistic graphical model (PGM) of student affinities towards STEM majors across time that is based on a modified Hidden Markov Model \cite{Koller2009}.

\subsubsection{Model overview}
The PGM (shown in Figure \ref{fig:hmm}) represents a student's term-by-term (quarter-by-quarter, in this case) STEM affinity as a series of $T$ dependent hidden states where $T$ is the number of terms that a student is at the university. We define ``STEM affinity'' as the calculated probability at any given term (\(t\)) of a student eventually graduating with a STEM degree. Each of the hidden states in the model ($X_t$) produces a plate of \(N\) observed states, with each observed state ($C_{ti}$) being a binary indication of whether or not a student took a specific course for all \(N\) possible courses (5,557 distinct courses in all). To simplify, all \(N\) observed states are assumed to be conditionally independent given the hidden state that produces them.

We use this particular model because it allows us to examine when a student's STEM affinity changes as they progress through their undergraduate academic career. The model's unsupervised nature allows us to describe certain latent properties of student transcript records (such as intent), which are not quantifiable for study using supervised ML approaches. This is advantageous because the calculated STEM affinities can serve as proxies for students' intentions with regards to pursuing STEM or non-STEM academic paths at each quarter. Thinking about these affinities as they relate to intentions captures this idea because a student's intentions influence which courses they take at each quarter just as the hidden states in the model produce the observed states. Though students' declared majors provide some indication of intent, students often begin pursuing alternate fields of study before formally declaring/changing majors. What's more, as discussed previously, most students enter the university with non-specific pre-major designations.
\input{dnbn}

\vspace{0.1cm}
\subsubsection{Data for graphical model}
\label{subsubsec:affinitydata}
To simplify, the data for this proof-of-concept model consists of a subset of the data described in Section \ref{subsec:data} --- it consists of all graduates (as defined in Section \ref{subsec:grads}) who entered UW in the 2004 calendar year. Limiting the data to a single cohort helps ensure that students have access to the same courses and majors, which are prone to change over time; limiting data to graduates provides a notion of ground truth with regards to whether or not a student eventually completed a STEM degree. This limited dataset consists of 3,960 graduates who completed any baccalaureate degree and their course history (137,348 transcript entries). The feature engineering process outlined in Section \ref{subsubsec:features} used for the supervised ML experiments was not used here. Features for the model consist only of a list of student courses taken. To further simplify, we only use binary indicators of whether or not students took a course (not their grades therein) and each student's entire transcript record is used to determine courses taken. As such, the dataset consists of 3,960 graduates and binary indicators of the courses they took for each academic term they were at UW; this dataset is henceforth referred to as the ``affinity dataset.'' About one quarter of the students in the affinity dataset were STEM graduates (23.5\%) while the remainder graduated with non-STEM degrees (76.5\%). The largest numbers of students graduated in 12 and 13 quarters (16\% and 13\% of students in the affinity dataset, respectively), which is consistent with taking about 3 academic quarters per year for 4 years.

\subsubsection{Model construction}
\label{subsubsec:construction}
We use the forward-backward algorithm \cite{rabiner1986introduction} to calculate term-by-term hidden state probabilities (STEM affinities) for each student based on their corresponding observed states (courses taken). To determine convergence, we set a threshold of 0.5 for the maximum log likelihood for the hidden state probabilities. A value of 0.5 for STEM affinity is used to define the threshold for switching between a STEM intention classification (STEM affinity $\geq$ 0.5) and a non-STEM intention classification (STEM affinity $<$ 0.5). The prior probability for the hidden states (\(X_0\)) is initialized as the proportion of students graduating with STEM degrees in the affinity dataset (0.235). The initial transition probability matrix is determined by a cost function which maximizes the number of students whose final-term STEM affinity matches the STEM categorization of their degree while also minimizing the number of students who switch between STEM and non-STEM trajectories after their first quarter. The emission probabilities for each of the \(N\) observed states (courses) are initialized as the proportion of STEM and non-STEM graduates in the affinity dataset who took each course. Laplace smoothing is applied to prevent zero values when calculating the emission probabilities.

\section{Results \& Discussion}
\subsection{Predicting STEM attrition}
\label{subsec:predRes}

\subsubsection{Machine learning experiments}

\begin{figure}[!]
\includegraphics[scale=1, trim={0 0.3cm 0 0}]{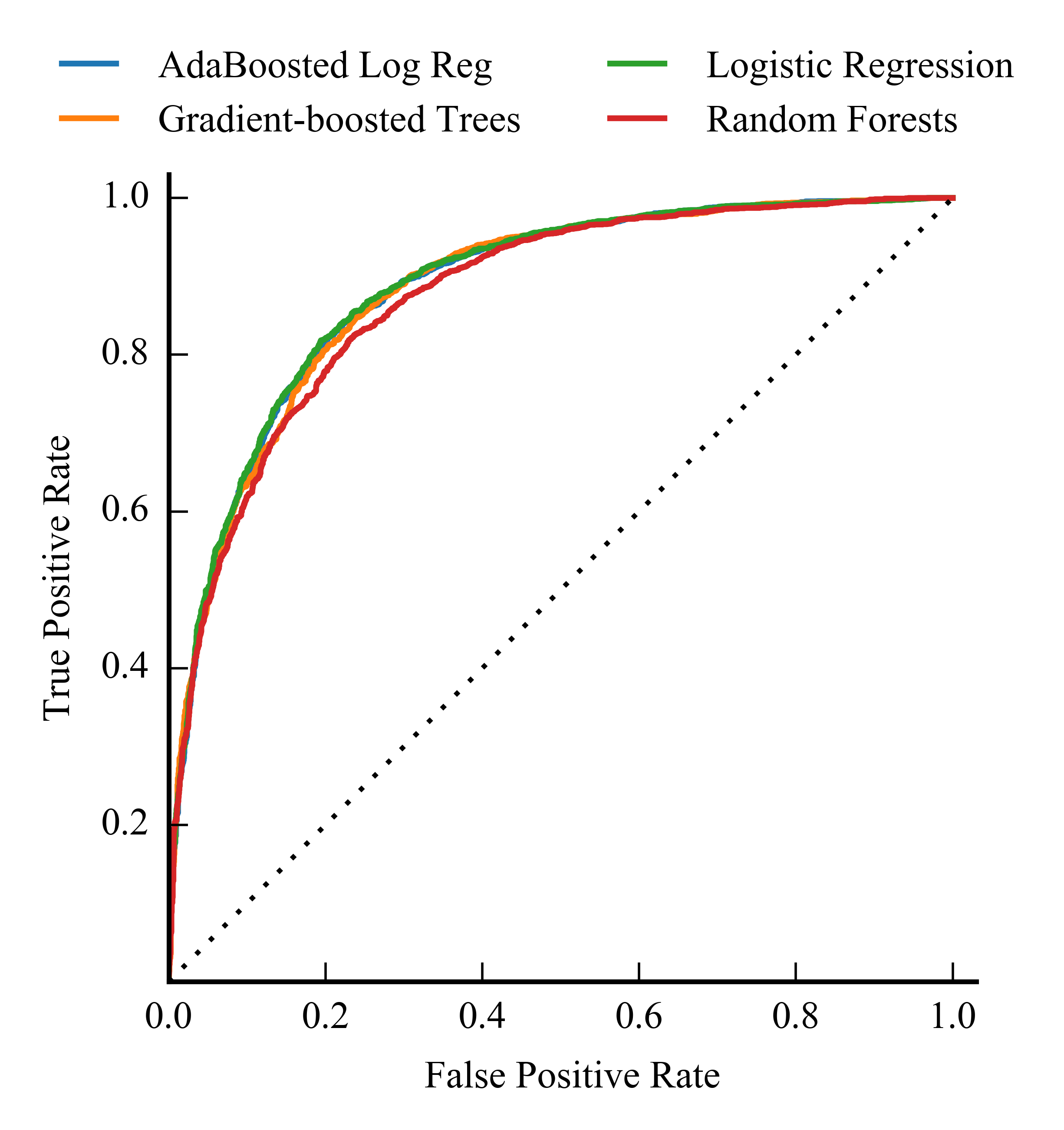}
\caption{Receiver operating characteristic curves for supervised ML models predicting STEM graduation.}
\label{fig:ROC}
\end{figure}

Receiver operating characteristic (ROC) curves for each of the four models are shown in Figure \ref{fig:ROC}. Prediction accuracies, the area under the curve of the ROC curves (AUROC), and F1 scores are shown in Table \ref{tab:preds}. In all, model performance was similar across all models (AUROCs between 0.87 and 0.89, F1 scores between 0.80 and 0.82), with logistic regression performing slightly better than the other models. Interestingly, boosting seemed to have no appreciable increase in predictive performance, as can be seen with the small performance difference between Ada-boosted logistic regression and logistic regression. Seeing logistic regression perform better than other models is in line with what our group has seen in the past when predicting dropout across all students \cite{Aulck2016}. As was the case with our previous work, we believe these results are strong given the limited amount of information fed into the models, as data was extracted from registrar records alone. That said, we also view these results as baselines to be improved upon in future work, particularly with respect to more extensive feature engineering and alternative modelling techniques, as discussed in Section \ref{sec:future}.

\begin{table}[h!]
  \caption{Model results for predicting STEM attrition}
  \label{tab:preds}
  \begin{tabular}{l|ccc}
    \toprule
    Model & Accuracy & AUROC & F1 Score\\
    \midrule
    Logistic Regression & 0.812 & 0.887 & 0.818\\ 
    Ada-boosted Log Reg & 0.806 & 0.885 & 0.811\\ 
    Gradient-boosted Trees & 0.806 & 0.885 & 0.810\\
    Random Forests & 0.792 & 0.874 & 0.799\\ 
  \bottomrule
\end{tabular}
\end{table}

\subsubsection{Individual predictors of STEM success}
To better understand individual factors related to STEM attrition, we also regressed the binary STEM graduation variable on a single feature at a time. When examining the 10 best individual predictors of STEM success as shown in Table \ref{tab:singles}, the importance of success in math classes to STEM students' educational pursuits becomes apparent. Five of the 10 most predictive features in isolation, including the four most predictive, were derived from student performance in math courses: 1) average GPA in math gatekeeper courses, 2) average percentile score in math gatekeeper courses, 3) average GPA in math courses, 4) average percentile score in math courses, and 9) average z-score in math classes. While it should be noted that these features are highly correlated, the fact that no other subject's grades were among the 10 most predictive features in isolation speaks to the weight of math courses in changing STEM-related student outcomes, particularly during students' first academic year. Though not specific to math, this idea of performance in introductory STEM classes relating to STEM success ties in with findings in previous work \cite{chen2015stem}.

\begin{table}[ht!]
  \caption{Best single predictors of STEM attrition}
  \label{tab:singles}
  \begin{tabular}{ll|ccc}
    \toprule
    & Feature & Accuracy & AUROC & F1 Score\\
    \midrule
    1. & MATH GK GPA & 0.704 & 0.764 & 0.733\\ 
    2. & MATH GK \%ILE & 0.703 & 0.762 & 0.732\\ 
    3. & MATH GPA & 0.698 & 0.758 & 0.736\\ 
    4. & MATH \%ILE & 0.697 & 0.756 & 0.734\\ 
    5. & AVG. Z-SCORE & 0.688 & 0.749 & 0.702\\ 
    6. & 100-LEVEL Z-SCORE & 0.686 & 0.751 & 0.701\\ 
    7. & \# CREDITS EARNED & 0.678 & 0.726 & 0.704\\ 
    8. & \# PASSED COURSES & 0.666 & 0.730 & 0.694\\ 
    9. & MATH Z-SCORE & 0.689 & 0.724 & 0.673\\ 
    10. & MAX. GRADE DIFF. & 0.670 & 0.723 & 0.692\\
  \bottomrule
\end{tabular}
\end{table}

\begin{figure*}[htb!]
\includegraphics[scale=1, trim={0 0.3cm 0 0}]{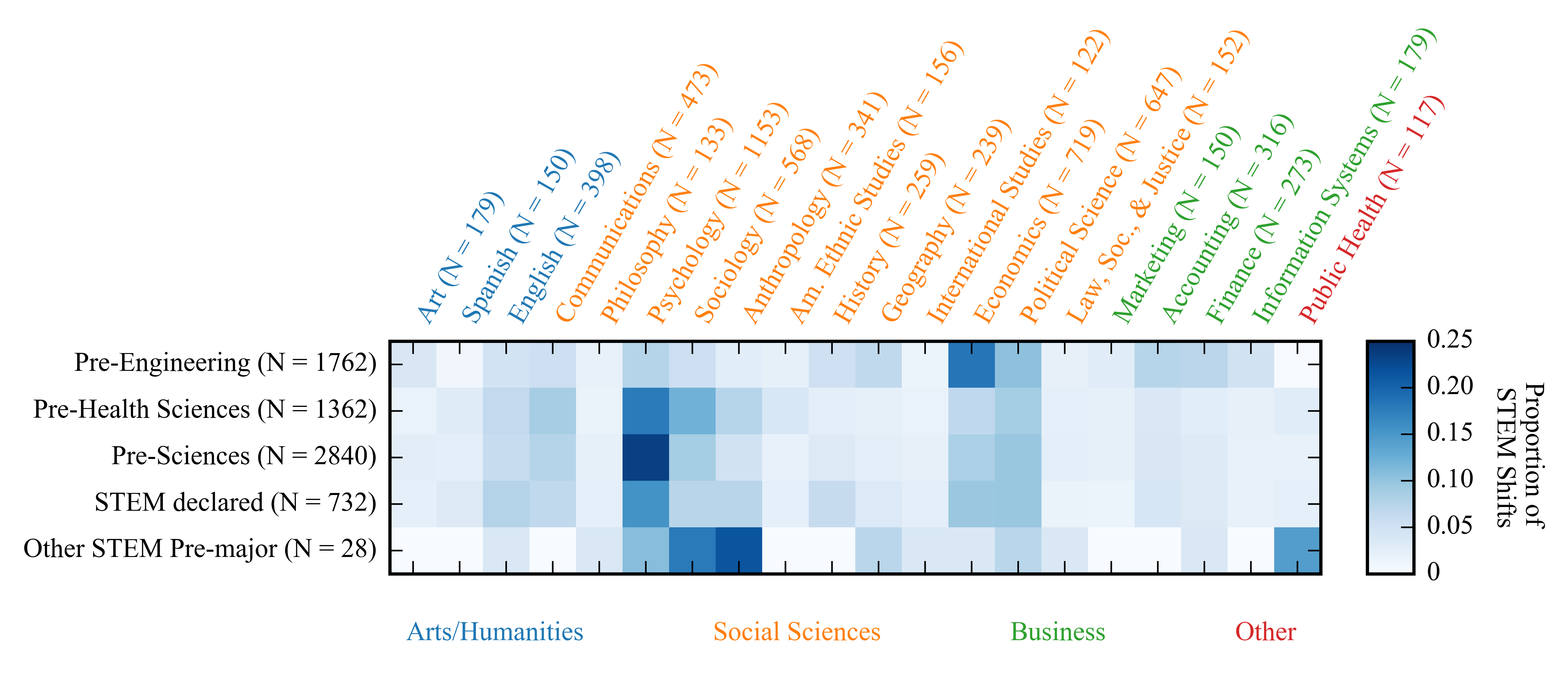}
\caption{Heatmap of non-STEM degrees obtained by STEM NCs. Rows indicate starting (STEM-related) major while columns indicate non-STEM degree earned. Counts are normalized across rows and numbers indicate counts of majors and degrees for each row/column. Only the 20 most frequently obtained degrees are shown and row counts only include students graduating with those degrees.}
\label{fig:heat}
\end{figure*}

In addition to math courses, students' average z-score in all courses and their average z-score in 100-level courses were also among the most highly predictive single features (5th and 6th, respectively). Interestingly, z-score calculations had higher predictive power than measures of GPA and percentile rank for these performance measures. General measures of student progress such as the number of credits earned by students as well as a count of the number of courses passed by students were also highly predictive (7th and 8th, respectively). The 10th most predictive feature was the average of the difference between students' grades in courses and the highest grade awarded in those same courses. Measures of performance in high school and standardized test scores, meanwhile, did not seem to have a large signal in predicting STEM attrition, in contrast to previous research \cite{rask2010attrition}. This could, however, be at least in part due to the breadth of high schools attended by students entering a large, public university and the lack of normalization of the grades therefrom. It could also be due to the highly competitive nature of many STEM majors.

\subsection{Understanding STEM transitions}
Better understanding STEM attrition also begs the question of where STEM students transition to upon leaving STEM fields. Of all STEM NCs in the dataset (12,220) about 60\% (7,303) ended up graduating with non-STEM degrees. Table \ref{tab:trans} presents the count and percentage of the non-STEM fields that these STEM NCs most frequently graduated from. It should be noted that due to changing major size/availability as well as the introduction of new majors across time, these counts were not normalized in any way. That said, the raw counts indicate that graduating STEM NCs most frequently graduated with degrees in psychology (13.45\% of graduating STEM NCs), which saw over 1.5 times as many students graduate than the field with the second-most graduates, economics (8.39\% of graduating STEM NCs). Further examining the list, the more popular majors tended to be in the social sciences with some from business (namely, accounting and finance) seen at the tail end of the top ten.

\begin{table}[ht!]
  \caption{Most frequently obtained degrees by STEM NCs}
  \label{tab:trans}
  \begin{tabular}{ll|ccc}
    \toprule
    & Major & Count & \shortstack{\% of Graduated\\STEM NCs}\\
    \midrule
    1. & Psychology & 1,153 & 13.45\% \\ 
    2. & Economics & 719 & 8.39\% \\ 
    3. & Political Science & 647 & 7.55\% \\ 
    4. & Sociology & 568 & 6.63\% \\ 
    5. & Communications & 473 & 5.18\% \\ 
    6. & English & 398 & 4.64\% \\ 
    7. & Anthropology & 341 & 3.98\% \\ 
    8. & Accounting & 316 & 3.69\% \\ 
    9. & Finance & 273 & 3.18\%\\ 
    10. & History & 259 & 3.02\% \\ 
  \bottomrule
\end{tabular}
\end{table}

To further examine these STEM transitions, Figure \ref{fig:heat} shows a heatmap  of graduating STEM NCs and their transitions from STEM-related fields to non-STEM fields. Our group has constructed similar heatmaps in the past to analyze student transitions across fields and believe them to be a way to better understand the multitude of paths students take towards degrees \cite{aulck2017attrition}. The rows indicate students' STEM majors/pre-majors declared during their first academic year, as described in Section \ref{subsec:data} when detailing the isolation of STEM students. The columns of the heatmap indicate the fields from which students graduated. Only the 20 most frequently obtained degrees are shown in the columns and the row counts only include students who graduated with these degrees. A few interesting patterns emerge when looking at these student transitions. Students with pre-engineering designations tended to favor going into economics rather than psychology, which was more popular with most other groups of graduating STEM NCs. Pre-engineering students also had a higher tendency to graduate from business-related fields (including accounting, finance, and information systems) than their graduating STEM NC peers. Students who declared STEM-related pre-majors that were not related to engineering, health sciences, or the physical sciences had more variability in their academic paths as they favored going into sociology, anthropology, and public health more than their peers. This group of students consisted of those who had declared a pre-major in a STEM field that had its own pre-major track and designation, a trend that has been growing at UW as additional competitive major groups are added. This group also had the smallest sample (n = 28) of the STEM student groups. In all, a majority of graduating STEM NCs earned degrees that were concentrated in the social sciences with little variation other than that noted above. It should be noted that this heatmap does not take into account transitions across or towards STEM majors as other studies have \cite{chen2015stem, bettinger2010or}, but only transitions away from STEM fields.

\subsection{Analyzing STEM affinities}

\begin{figure}[h!]
\includegraphics[scale=1, trim={0cm 0.3cm 0 0}]{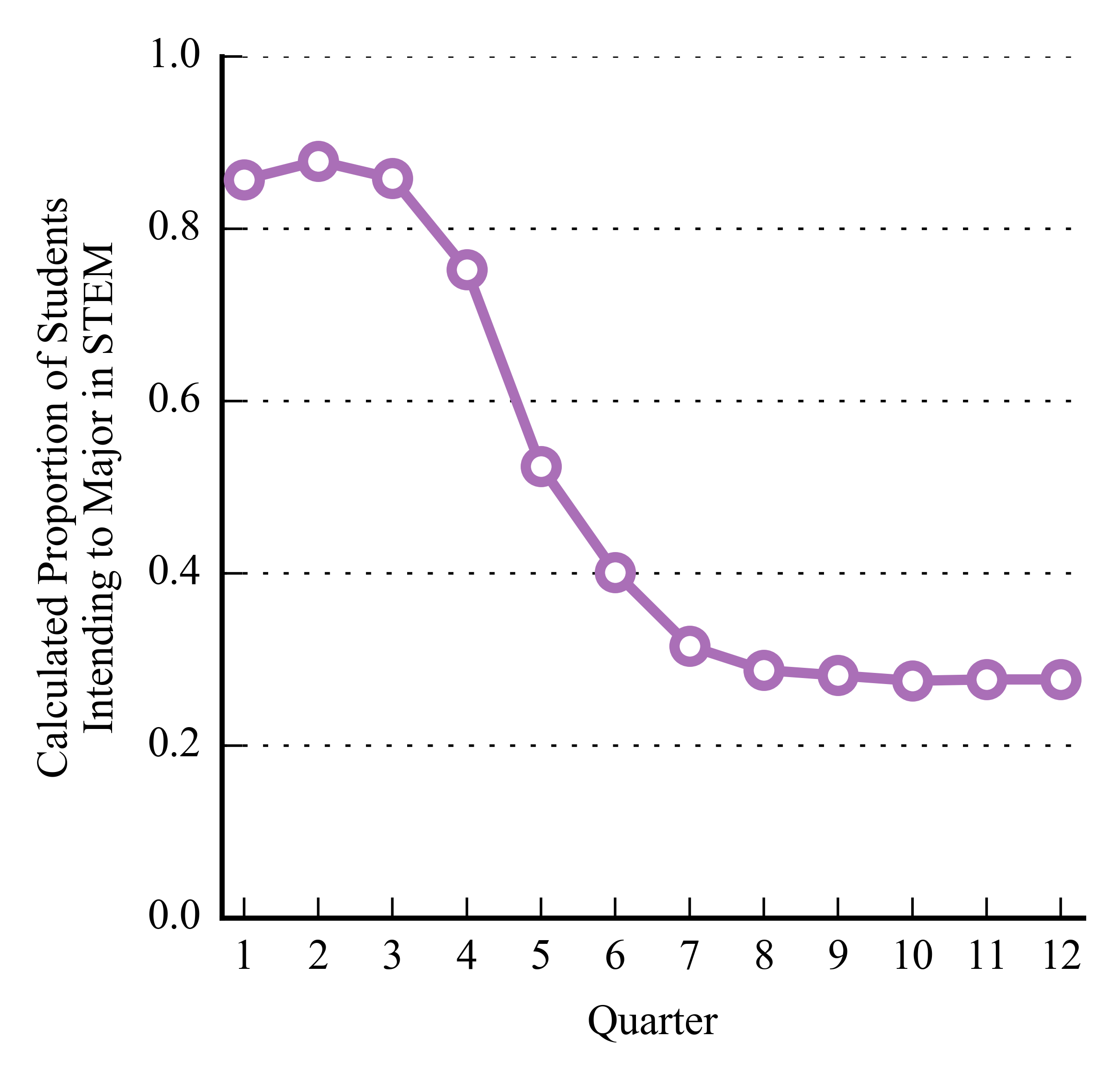}
\caption{Calculated proportion of students intending to major in STEM across time for all students graduating in exactly 12 quarters. Note the large decrease in prospective STEM students during their 2nd year on campus.}
\label{fig:hmmResult}
\end{figure}

To validate the performance of our graphical model, we compared whether students' calculated designation of STEM or non-STEM for their final quarter (based on a STEM affinity threshold of 0.5) with whether or not they were a STEM graduate. Here, the STEM affinity (intention) for the last quarter can be checked because we know what the students ultimately graduated with; for all previous quarters, however, it cannot as there is no measure of students' STEM intention. Based on this, the accuracy, recall, and precision of the model were 0.896, 0.999, and 0.670, respectively. 

Figure \ref{fig:hmmResult} shows the estimated proportion of STEM-labelled students across time for students that graduate in exactly 12 quarters. This subset of the data (i.e. the 643 students graduating in exactly 12 quarters) was chosen because, as mentioned in Section \ref{subsubsec:affinitydata}, the largest proportion of students in the affinity dataset graduated in 12 quarters. Additionally, 12 quarters also represents the typical 4-year undergraduate time-to-completion. As shown, the proportion of these students with STEM trajectories begins relatively high in the first three quarters and sharply drops thereafter before stabilizing around the 7th quarter, which typically corresponds to the start of students' 3rd academic year. The sharp decline, meanwhile, occurs during most students' 2nd year, just prior to when most students are required to formalize their major decisions. It is during this pivotal time that one would expect STEM attrition to occur. The initial spike in STEM-interested students, we believe, can at least in part be attributed to noise in the model, which is discussed in greater detail below.

\begin{figure}[h!]
\includegraphics[scale=1, trim={0cm 0.3cm 0 0}]{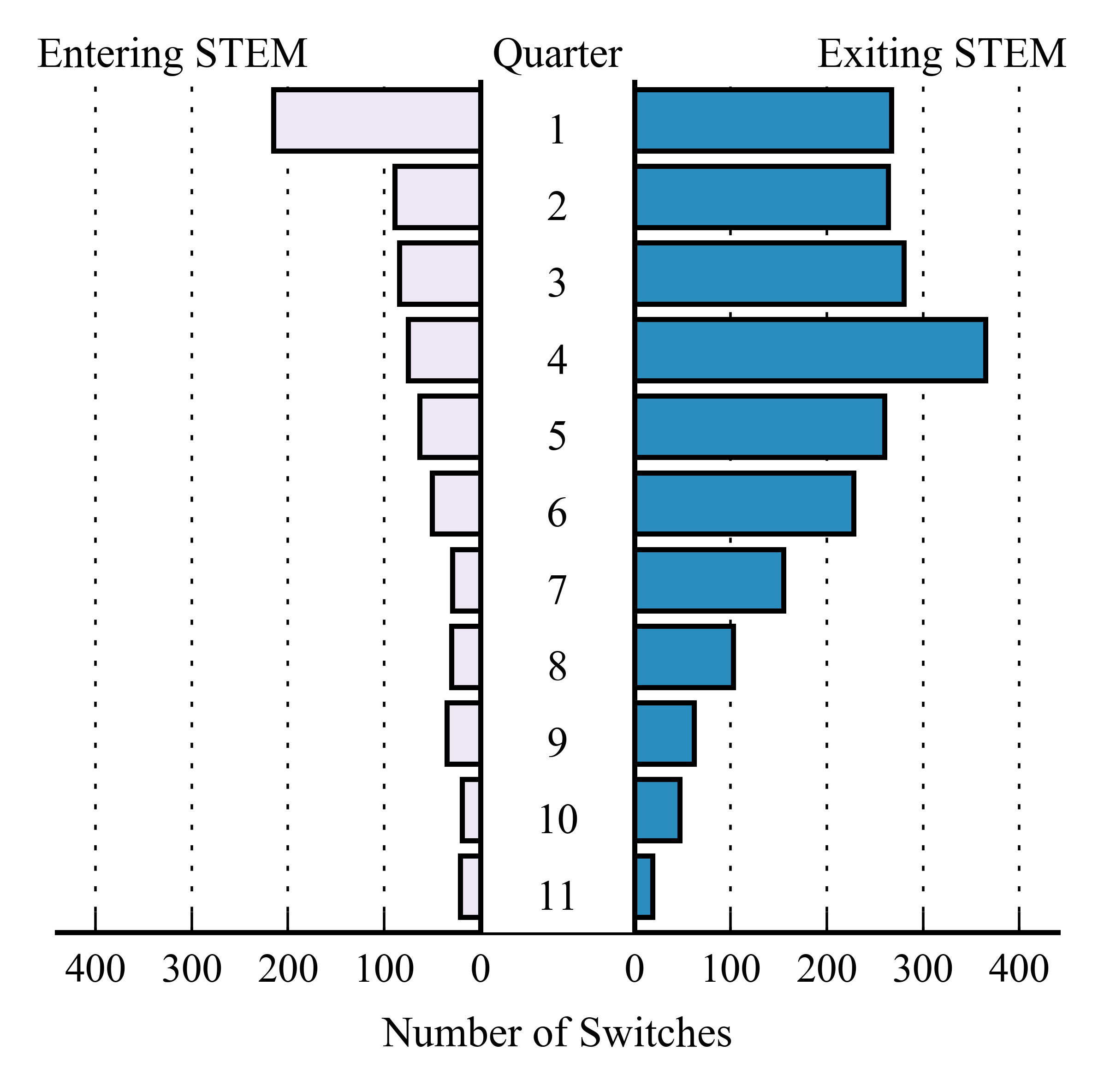}
\caption{Calculated counts of students switching into (left) or out of (right) STEM trajectories by quarter at the university. Calculations are based on STEM affinities for each student at the end of each quarter. Only the first 11 quarters are shown as a majority of students graduate within 12 quarters.}
\label{fig:intentions}
\end{figure}

\begin{figure*}[ht!]
\includegraphics[scale=1, trim={0 0.1cm 0 0}]{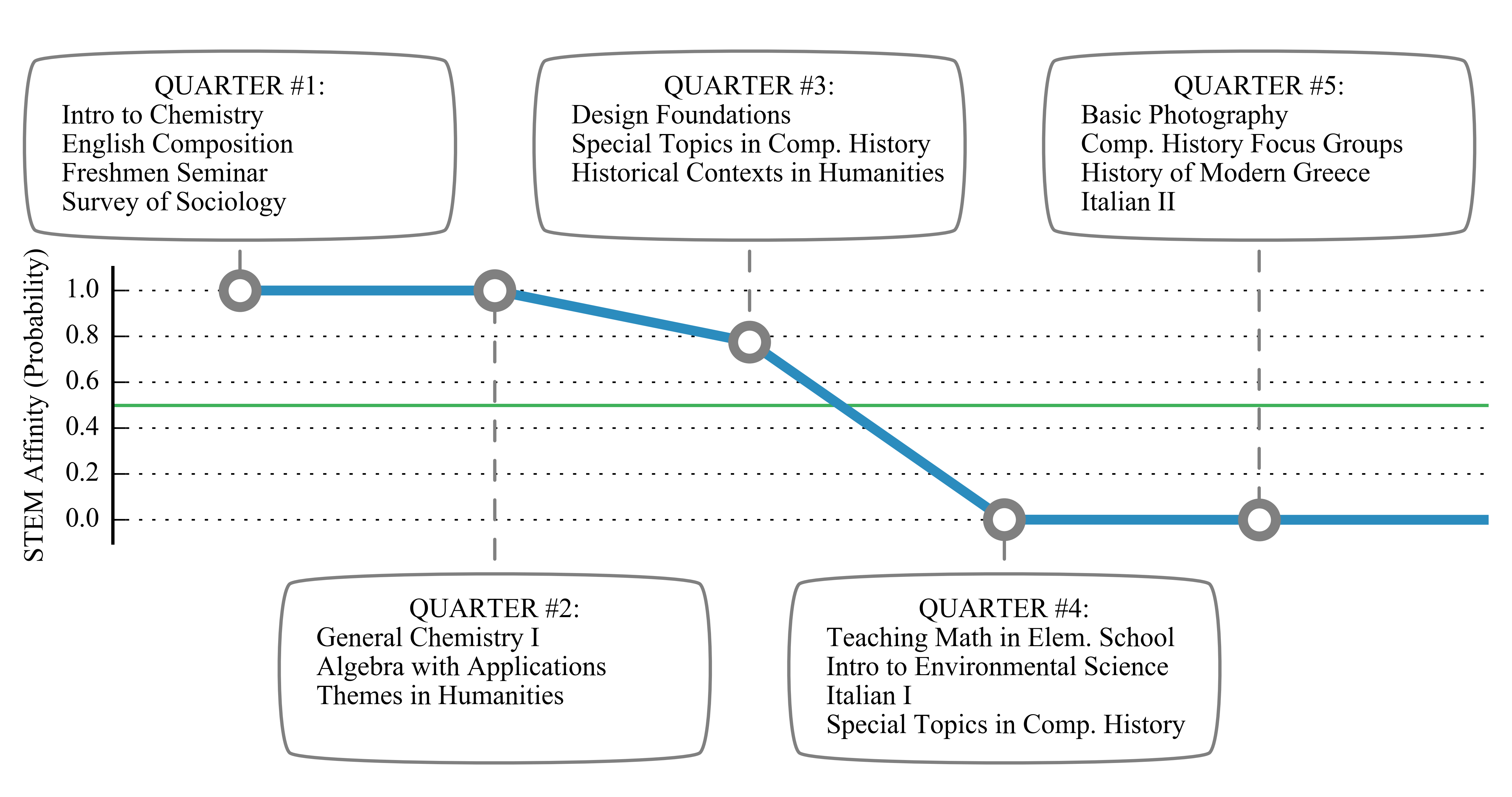}
\caption{STEM affinity across time for a single student in the dataset as a case study. The affinities are presented for each quarter and the corresponding courses that the student took in each quarter are shown. The green line indicates the 0.5 threshold for switching from a STEM to non-STEM trajectory. Note how the first two quarters see the student taking introductory STEM courses while the latter quarters are more focused on the humanities and social sciences.}
\label{fig:hmm_case}
\end{figure*}
A switch in a student's STEM intention classification occurs when the student's intention classification from a given quarter differs from their label in their previous quarter. Analytically, this is determined by finding two adjacent quarters where STEM affinity scores cross the 0.5 threshold that determines intention classification. Figure \ref{fig:intentions} shows the number of switches per quarter across the entirety of the affinity dataset on a quarter-by-quarter basis. As expected, students switch STEM intentions frequently during the beginning of their academic careers. This aligns with the intuition that student intentions are more volatile at the start of their schooling due to a number of factors. First, general education requirements introduce noise into the model as students take a wide range of courses simply to fulfill requirements early on. Second, students often undergo an exploratory phase in their education to better scope their interests upon entering college. Third, student intentions correct to their ``true'' values following their initial assignment based on our estimated prior. Figure \ref{fig:intentions} also shows a spike in STEM attrition occurring after the 4th quarter, which aligns with what is seen in Figure \ref{fig:hmmResult}. In all, most quarters after students' 1st see a much higher rate of students switching away from STEM intentions than students orienting towards STEM pursuits. 

In Figure \ref{fig:hmm_case}, we show a single student's STEM affinity scores over time along with the courses they took as a case study. This specific student provides an example of a student shifting from a STEM trajectory to a non-STEM trajectory. Examining the graph, the student's affinity scores in the first 2 quarters strongly suggest a heavy inclination towards a STEM major. The corresponding courses the student took during these two quarters provides context to this - about half the students' courses (3 of 7) are in pre-STEM topics as the student takes introductory math and chemistry courses. Meanwhile, the other courses the student took during this period were very generalized electives which did not heavily weigh in the calculations of their STEM affinity. At the end of the student's third quarter, however, their STEM affinity decreased substantially and their fourth quarter courses pushed their STEM affinity beyond the transition point (marked by the green line in the Figure). The courses taken in these two quarters are strongly associated with non-STEM fields, as the student began focusing on comparative history, foreign languages, and the social sciences at large while not taking any STEM-related courses. Their courses thereafter align with their low STEM affinity score as the student continues to take courses in the humanities and social sciences. The student remains on this non-STEM trajectory after their early transition and eventually graduates with a non-STEM degree. Observations such as these are interpreted as the student's intended major switching from a STEM track to a non-STEM track. This student's STEM affinity plot also provides examples of the transition points we are interested in further identifying and understanding, particularly around the third quarter.

In all, we believe our initial attempts at modelling STEM affinities and, by proxy, student STEM intentions provides a promising first step in identifying when students change their academic interests. That said, we also believe there are numerous ways to expand this approach to better model student trajectories, as discussed further in Section \ref{sec:future}.

\section{Future Directions}
\label{sec:future}
We believe this work provides many potential avenues for expansion and further exploration. For the supervised predictions of STEM attrition, more expansive feature engineering should improve model performance. One idea we intend to explore relates to the fact that most students who enter UW do so with pre-major designations of some sort (about 81\% of students, as calculated from our dataset) and then declare a major in their 2nd or 3rd year on campus. If a student is a pre-major, it is difficult to determine which specific course of study they are interested in pursuing though there may be some broad sense (i.e. pre-engineering, pre-health sciences, etc.). However, requirements for entry and the level of competition for limited spots\footnote{See: \url{http://bit.ly/2kVjV1z}} vary greatly across majors and having a better understanding of how a student is faring compared to peers intending to enter the same field should give a strong indication of how likely they are to succeed in following their current academic trajectory. To better understand this, we intend to use students' course history to estimate their major(s) of interest (perhaps using expanded affinity scores similar to as we do in this work). From there, we can compare each student to weighted composites of students who entered the majors of interest for the student in order to get a sense of how the student is faring as a pre-major. Additionally, we are also interested in additional technical improvements to our models, including the use of neural networks to improve predictions and believe predictive power can be improved using alternative modelling techniques.

To expand the work on analyzing STEM affinities, we plan to further develop our model to better understand when students transition away from STEM fields. Currently, there is no way to validate the STEM affinity scores against a student's actual STEM intentions at each time period. While we tuned the model according to the cost function described in Section \ref{subsubsec:construction}, it only examines STEM affinities at the beginning and end of their transcript. We hope to explore options to include more comprehensive heuristics that will reflect the long-term behavior of our model. In addition, the current state space of the model captures relatively high-level STEM and non-STEM probabilities that could lead to potential loss of information during transitions. As mentioned above, we are interested in expanding the state space from STEM and non-STEM labels to each to individual major. Additionally, we only use a small subset of the data available to us in modelling the students in this work. We believe adjusting our model assumptions/parameters such that data from multiple cohorts can be used should provide a more comprehensive look at STEM affinities. Experimenting with different objective functions for optimizing the model may also be further explored, particularly with improving model precision.

Another potential idea to better identify the transition points that the graphical model attempts to elucidate is to add properties such as ``affinity inertia,'' which would aggregate affinity scores from multiple previous quarters, thereby reducing the impact of a single transcript entry in our model. We believe this affinity inertia could reduce the noise generated in our model from general education requirements as shown in some of the results shown here. Upon better identifying transition points, we could use supervised machine learning methods to understand what additional features from student transcripts predict changes in student STEM affinity and, by proxy, student STEM intentions. These features could include grades, demographics, and course evaluation information. Additionally, we are also actively working with UW to better collect and curate student intention information for entering students, which we believe will provide another information-rich data source for our models. 


Ultimately, we believe the entirety of this work has many applications in advising at university campuses. Identifying students who are likely to change their academic interests provides a wealth of opportunity to develop interventions for identifying struggling students and preventing them from dropping out. Furthermore, the proof-of-concept graphical model we present provides an idea of not only \textit{who} will attrite from STEM fields but also \textit{when} they may be inclined to do so. This understanding of when can be crucial to providing students assistance in a timely manner.

\section{Conclusions}
In this work, we present results for predicting and understanding STEM attrition from a large, heterogeneous dataset of students using only data mined from university registrar databases. The predictions yielded promising results with about a 30 percentage point increase in accuracy over the baseline proportions. Analyzing individual predictors of STEM attrition, meanwhile, gave support to the idea that math classes and entry STEM classes play crucial roles in students' pursuit of STEM degrees.

We also found that students shifting out of STEM follow varying paths to completion, particularly pre-engineering students. Compared to their peers, we found pre-engineering student to have a greater propensity to graduate in economics and business-related fields. Meanwhile, across all students who switched away from STEM pursuits, psychology was by far the most popular degree earned, not adjusting for the size of the major.

Our probabilistic graphical model provided a proof-of-concept approach to modelling term-by-term student intentions as they progress in their undergraduate academics. Using the affinity scores we calculated, we were able to outline points at which students tend to transition away from STEM fields. Namely, a large amount of switching occurs early in students' academic careers or near the time when they are required to declare a major (typically their 2nd year, in our dataset). After this, student STEM intentions were relatively stable. 


Our next steps for this work include improving our supervised machine learning models, expanding the scope of our probabilistic graphical models, and working with university administrators to develop interventions. For more information on our research, visit \url{www.coursector.org}.


\begin{acks}
The authors would like to thank the University of Washington Registrar and the University of Washington Office of Planning and Budgeting for their help in obtaining the data. The authors would also like to thank Dev Nambi, Noah Smith, and Steve Tanimoto of the University of Washington as well as Joshua Blumenstock of the University of California-Berkeley for their support of the work and their insights into the models.
\end{acks}

%% file: dnbn.tex
\tikzstyle{state}=[shape=circle,text=black,draw=blue!50, minimum size=10]
\tikzstyle{lightedge}=[<-,dotted]
\tikzstyle{latent}=[state,thick]
\tikzstyle{obs}=[state, fill=orange!20]
\tikzstyle{mainedge}=[<-,thick]

\begin{figure}[h]
\begin{tikzpicture}[]
\node               at (0,4.3) {$t=0$};
\node[latent] (x0) at (0,3.6) {$X_0$};
\node               at (1.9,4.3) {$t=1$};
\node[latent] (x1) at (1.9,3.6) {$X_1$}
    edge [mainedge] (x0);
\node[obs] (c1) at (1.9,2.1) {$C_{1i}$}
    edge [lightedge] (x1);
\node               at (3.8,4.3) {$t=2$};
\node[latent] (x2) at (3.8,3.6) {$X_2$}
    edge [mainedge] (x1);
\node[obs] (c2) at (3.8,2.1) {$C_{2i}$}
    edge [lightedge] (x2);
\node               at (5.7,4.3) {$t=T$};
\node[latent] (xT) at (5.7,3.6) {$X_T$};
\node[obs] (cT) at (5.7,2.1) {$C_{T}$}
    edge [lightedge] (xT);
\path (xT) -- node[auto=false]{\ldots} (x2);

\plate [inner sep=.3cm,xshift=.05cm, yshift=.2cm] {plate1} {(c1)} {courses $i$};
\plate [inner sep=.3cm,xshift=.05cm, yshift=.2cm] {plate2} {(c2)} {courses $i$};
\plate [inner sep=.3cm,xshift=.05cm, yshift=.2cm] {plate3} {(cT)} {courses $i$};

\end{tikzpicture}
\caption{Illustration of DNBN used to estimate STEM affinities at every academic term. Solid arrows indicate transitions between states and dashed arrows indicate emissions. Plates index over all possible courses and are shown as boxes. $X_0$ is the prior probability distribution.}
\label{fig:hmm}
\end{figure}